\documentclass[journal]{IEEEtran}
\ifCLASSINFOpdf
\else
\fi
\usepackage{url}

\usepackage[hidelinks]{hyperref}
\usepackage{graphicx}
\usepackage{enumitem}
\usepackage{xcolor}
\usepackage{booktabs}
\usepackage[caption=false,font=footnotesize]{subfig}
\usepackage{orcidlink}
\usepackage{booktabs}
\usepackage{amssymb}
\usepackage{threeparttable}
\usepackage{multirow}

\newcommand{\name}{NetLLMeval}

\usepackage{colortbl}
\usepackage{framed}
\usepackage{scalerel}
\usepackage{changepage}

\definecolor{formalshade}{rgb}{0.9,0.95,1}

\definecolor{darkblue}{rgb}{0.6,0.75,0.95}

\newenvironment{formal}{%
  \MakeFramed{\advance\hsize-\width\FrameRestore}%
  \noindent\hspace{-4.55pt}
  \begin{adjustwidth}{}{7pt}%
}
{%
  \end{adjustwidth}\endMakeFramed%
}

\definecolor{formalshade2}{rgb}{1,0.85,0.85}
\definecolor{darkred}{rgb}{0.6,0.0,0.0} 

\begin{document}
\bstctlcite{IEEEctl:options}
%
\title{Toward Agentic SysAdmin: Rethinking System Administration with AI Agents}
%
%
%

\author{
Gianmaria~Frigo~\orcidlink{0009-0009-3030-8982},
Davide~Saladino~\orcidlink{0009-0000-4865-1492},
Alberto~Castagnaro~\orcidlink{0009-0008-1809-2253},
Francesco~Marchiori~\orcidlink{0000-0001-5282-0965},
Denis~Donadel~\orcidlink{0000-0002-7050-9369},
Luca~Pajola~\orcidlink{0000-0002-6749-6608},
Mauro~Conti~\orcidlink{0000-0002-3612-1934},~\IEEEmembership{Fellow, IEEE}
\thanks{G. Frigo, D. Saladino, F. Marchiori, and M. Conti are with the University of Padova, Italy. M. Conti is also with Orebro University, Sweden. A. Castagnaro and L. Pajola are with Spritz Matter SRL, Italy. D. Donadel is with the University of Verona, Italy. Emails: \{gianmaria.frigo, davide.saladino, francesco.marchiori.4\}@studenti.unipd.it, mauro.conti@unipd.it, \{luca.pajola, alberto.castagnaro\}@spritzmatter.it, denis.donadel@univr.it.}
}

%
%

\markboth{IEEE Transactions on Network and Service Management
}{}%
%



\maketitle

\begin{abstract}
The growing complexity of computer networks, driven by cloud-native architectures, heterogeneous devices, and distributed systems, places increasing pressure on network administrators who must simultaneously manage configuration, troubleshooting, and security under tight operational constraints. Large Language Models (LLMs) have emerged as a promising tool to assist and partially automate these tasks, yet their systematic evaluation in networking scenarios remains an open challenge. Existing benchmarks rely on static reference outputs or manual expert validation, neither of which scales to the diversity of real-world network states or to the variety of orchestration strategies---from monolithic prompting to fully agentic pipelines---through which LLMs are increasingly deployed.

In this paper, we present \name{}, a framework for automatically evaluating LLM-based systems on network administration tasks by leveraging live network emulation to derive ground truth without human intervention. Through a full-factorial study of $24{,}000$ runs spanning 10 foundation models, 4 solver architectures, 10 task types, and 6 network topologies of increasing complexity, we show that solver design has a great impact on accuracy---lifting a 14B open-weight model from $0.43$ to $0.88$ correctness---and that such locally-deployable models can match trillion-parameter frontier systems under the right configuration. \name{} is released open-source to support reproducible benchmarking of future models and solver designs.

\end{abstract}


%
\IEEEpeerreviewmaketitle

\section{Introduction}

Computer networks are becoming increasingly complex, driven by the proliferation of distributed systems, cloud-native architectures, and heterogeneous devices~\cite{lindsay2021evolution}. This growing complexity makes network management and operation progressively more challenging~\cite{benson2009unraveling}. System administrators must simultaneously handle a variety of tasks, including configuring devices, troubleshooting connectivity issues, managing incidents, and responding to security alerts. These activities require both domain expertise and timely decision-making, often under significant operational pressure.

Recent advances in Artificial Intelligence (AI), particularly Large Language Models (LLMs), offer a promising avenue to support and partially automate network management tasks~\cite{wang2017machine, liu2024llmnetworkingworkflow}. LLMs are deep learning architectures trained on large-scale textual data, capable of understanding context, reasoning over inputs, and generating structured, human-like responses~\cite{raiaan2024review}. These capabilities make them suitable for assisting system administrators in diagnosing and resolving network issues. Prior work has demonstrated the potential of LLMs as effective assistants in networking scenarios~\cite{donadel2024can, wang2024netconfeval}.

However, evaluating the capabilities of LLMs in this domain remains a significant challenge. A central issue is the lack of reliable and objective ground truth for network-related tasks. Unlike other domains where answers can be easily verified, network troubleshooting and management often depend on the dynamic state of a specific system. As a result, validating LLM-generated responses typically requires expert manual inspection, which is time-consuming, error-prone, and not scalable. 
Execution-grounded evaluation, where correctness is derived automatically from a live environment rather than from static references or manual scoring, has begun to address this in adjacent domains~\cite{chen2025aiopslab,zhou2026netarena,wang2025network}. 

To effectively assess LLM performance, it is therefore necessary to develop mechanisms that can generate precise, programmatic ground truth for networking tasks. Such a capability would enable objective, reproducible benchmarking by providing definitive answers to network-related queries, automating evaluation, and enabling large-scale experimentation. Crucially, it would also make it possible to compare, on equal footing, the different ways an LLM can be deployed to solve a networking task.

This last point matters because a language model is rarely used in isolation. The same underlying model can be embedded in very different system designs: it may be queried directly, augmented with external tools that let it inspect the network, or wrapped in a multi-step procedure that lets it gather evidence and revise its answer before responding. We refer to these designs collectively as \textit{solver architectures}, and detail them in Section~\ref{sec:exp}.
The choice of architecture is not a minor implementation detail. A well-designed architecture may let a smaller foundational model match a much larger one, allow the use of locally deployed open-weight models that keep sensitive network data on-premise, reduce the number of tokens, and therefore the cost, required to reach an answer, and improve accuracy overall. Yet the size of this effect and whether it holds for weaker as well as stronger models are not well understood. Establishing this rigorously requires exactly the objective, automatically generated ground truth described above, applied uniformly across many models and architecture combinations.

\emph{Contributions.} In this paper, we propose \name{}, a framework for the automatic evaluation of LLM-based pipelines in networking scenarios. \name{} leverages network emulation to create controlled, reproducible environments in which a query is posed against a concrete network, the LLM-based system under test produces an answer, and that answer is verified automatically against the true state of the emulated network. Because the network is emulated, its true state is known by construction, yielding reliable ground truth without expert intervention and providing a scalable foundation for benchmarking both standalone LLMs and more complex agentic solutions. Using \name{} as our measurement instrument, we conduct a systematic comparison of model and solver-architecture combinations for network reasoning, structured around the following research questions:
\begin{enumerate}[label=\textbf{RQ\arabic*}]
    \item \textbf{Baseline capability.} What is the 
    capability of LLMs in network reasoning tasks, and how does performance vary with task difficulty and network topology complexity?
    \item \textbf{Effect of solver architecture.} How does  
    solver architecture affect accuracy, and does the benefit of more sophisticated solvers depend on the capability of the underlying model?
    \item \textbf{Local versus frontier models.} 
    Can locally deployable open-weight models, when paired with appropriate solver architectures, become a viable on-premises alternative to frontier-scale API models for network reasoning, matching their accuracy while preserving data privacy and reducing costs?
    \item \textbf{Cost–accuracy trade-offs.} What are the trade-offs between computational cost (token consumption and latency) and accuracy across model–solver configurations, and which configurations offer the best efficiency–accuracy balance?
    \item \textbf{Failure characterization.} What are the characteristic failure modes of different model–solver pairs, and what do execution traces reveal about the root causes of incorrect or invalid outputs?
\end{enumerate}
 
Our contributions can be summarized as follows:
\begin{enumerate}
    \item We design \name{}, a framework for scalable and automated evaluation of LLMs on network administration tasks, using live network emulation to derive ground truth without manual validation.
    \item We implement four solver architectures---Bulk, Bulk+ReAct, Guided Retrieval Agent, and Planner Agent---spanning from monolithic prompting to fully agentic pipelines.
    \item We employed \name{} to benchmark our solvers across 10 tasks and 6 labs, totaling more than $24{,}000$ runs across 10 foundation models, showing that locally deployable open-weight models can match frontier-scale API models under the right solver configuration.
    \item We open-source \name{} to support reproducible benchmarking of novel models and solver designs in networking\footnote{\url{https://github.com/pajola/agentic-sysadmin/}}.
\end{enumerate}
\section{Related Work}\label{sec:related}

\emph{LLMs for networking.} The application of LLMs to networking has attracted substantial attention~\cite{liu2024llmnetworkingworkflow, hong2025comprehensive}. NetLLM~\cite{wu2024netllm} proposes a unified adaptation framework that fine-tunes LLMs on networking problems, establishing that general-purpose models benefit from domain adaptation. Related efforts have targeted 5G network analysis~\cite{kan2024mobile}, security anomaly detection~\cite{entezami2025novel}, and protocol bug identification~\cite{zheng2025llm}, indicating broad applicability across the networking stack.

\emph{Evaluation of LLMs in networking and operations.} Systematic evaluation of LLMs on operational tasks is an active and rapidly evolving area. Our earlier study~\cite{donadel2024can} provided preliminary evidence that general-purpose LLMs can assist network administration but relied on manual validation and a limited question set. NetConfEval~\cite{wang2024netconfeval} benchmarks LLMs on network configuration tasks against static reference outputs, without closed-loop interaction with a live environment and without agent-based pipelines. The most closely related concurrent works are NetArena~\cite{zhou2026netarena} and NIKA~\cite{wang2025network}, both of which use network emulators to evaluate LLM agents. NetArena generates queries stochastically, mainly asking the models to fix a networking issue through sequential manipulations of the network that agents can do automatically.  
While this approach is interesting, the low accuracy suggests limited applicability in real-world scenarios involving live networks.
Instead, NIKA focuses specifically on network troubleshooting and incident resolution across five network scenario types. Beyond networking, AIOpsLab~\cite{chen2025aiopslab} evaluates LLM agents on cloud operations tasks in live microservice environments, establishing a complementary benchmark at the DevOps layer. Our framework differs from all of these in its explicit focus on comparing solver architectures, on providing configuration files directly on the prompt, on our evaluation of local versus frontier models at scale, and in its trajectory-level failure analysis, which none of the above provide.







\section{\name{} Architecture}

\begin{figure}
    \centering
    \includegraphics[width=.7\linewidth]{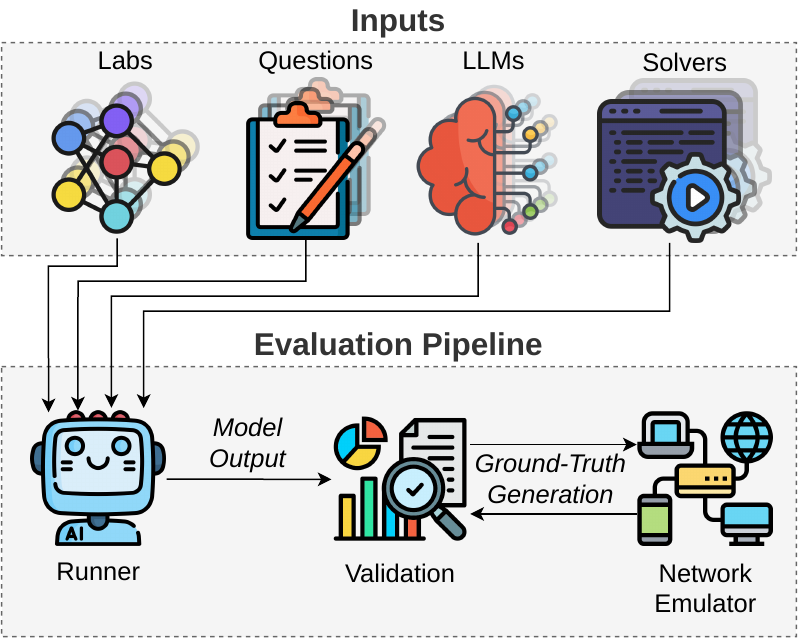}
    \caption{\name{} high-level architecture.}
    \label{fig:architecture}
\end{figure}

The \name{} framework is designed as a set of modules. The evaluation pipeline contains a runner, an analysis script, and a network emulator. It takes four components per round: a foundational LLM, a solver, a lab, and a question. For each round, the analysis pipeline verifies, directly in the network emulator, whether the solver's answer is correct.  

\subsection{Inputs}
\emph{\textbf{Labs}} define the network environments in which evaluation takes place. Each lab consists of a fully specified, executable network topology that includes nodes, links, configurations, routing protocols, and running services. 
Labs are 
instantiated within a network emulator and represent the source of ground truth for all tasks. By varying structural properties such as topology size, connectivity patterns, protocol diversity, and service deployment, labs provide a controlled measure of task difficulty and enable evaluation across heterogeneous, progressively complex scenarios.

\emph{\textbf{Questions}} formalize the tasks posed to solvers and are implemented as modular, executable components. Each question is defined by three elements: (i) a natural language prompt presented to the solver, (ii) a structured output specification describing the expected answer format (e.g., via a schema or data model), and (iii) a ground-truth generation function that programmatically derives the correct answer by interacting with the live emulated network. This design ensures that each task is both human-interpretable and automatically verifiable, enabling objective, reproducible, and scalable evaluation.

\emph{\textbf{LLMs}} are the foundational reasoning engines used in solvers. While tightly coupled with solver behavior, they are treated as a distinct dimension to enable controlled comparisons across models with different capabilities, sizes, and deployment settings (e.g., local vs. remote). This separation enables the framework to isolate the contribution of the underlying model from that of the orchestration strategy, supporting fair and reproducible benchmarking.

\emph{\textbf{Solvers}} define the orchestration logic for answering given network questions. Rather than being limited to a single prompt-response interaction, a solver encapsulates the full reasoning pipeline, including prompt design, intermediate reasoning steps, tool usage, and interaction patterns with the environment. This includes, but is not limited to, direct prompting strategies, multi-step reasoning frameworks (e.g., ReAct-style pipelines~\cite{yao2023reactsynergizingreasoningacting}), and multi-agent systems. By abstracting these design choices, \name{} enables the evaluation of how different reasoning and orchestration strategies affect performance on networking tasks, making it easier to develop and evaluate new solutions.

\subsection{Evaluation Pipeline}

The evaluation pipeline is designed to provide an objective, reproducible, and scalable assessment of responses in network administration tasks. At its core, the framework relies on programmatic \textit{ground-truth generation}, where correct answers are computed independently of the models by directly interacting with the network emulation environment. This removes human bias and ensures consistency across all experiments.

For each question, a dedicated procedure derives the expected answer by querying the emulated network. Depending on the task, this may involve analyzing network topology, parsing device configurations, inspecting running services, or verifying connectivity and reachability between hosts. By grounding all answers in the actual network state, the framework ensures that the evaluation reflects realistic and scenario-specific conditions. 

The validation process is fully automated and based on structured comparisons between the agent's outputs and the ground truth. Responses are normalized to comparable representations and evaluated for semantic correctness, allowing minor formatting variations while maintaining strict validation of the underlying content.
As our solution does not allow agents to directly interact with the network during question answering~\cite{zhou2026netarena}, except to retrieve configuration files, it can be deployed in real-world and safety-critical environments, where actions on the network should be supervised by an operator.


During execution, the system collects detailed information for each run, including accuracy metrics, execution time, error traces, and structured differences between predicted and expected outputs. This comprehensive data collection enables in-depth analysis of model behavior and failure modes.


\section{Experiments}\label{sec:exp}

\subsection{Questions}

A standardized set of network administration questions was applied to all labs to systematically evaluate the capabilities of the considered solvers and models. The questions are designed to cover a broad spectrum of network management tasks, including topology discovery, interface configuration analysis, service identification, and connectivity reasoning.

\begin{table}[tbh]
\addtolength{\tabcolsep}{-0.4em}
\caption{Questions employed in the pipeline. Full questions without cuts [...] are reported on our GitHub repository$^1$. \textbf{C}omplexity: \textbf{E}asy, \textbf{M}edium, \textbf{H}ard.}
\label{tab:questions}
\resizebox{\columnwidth}{!}{%
\begin{tabular}{ccl}
\toprule
\textbf{ID}  & \textbf{C} & \textbf{Question}     \\                                                                                                                               
\midrule
T1  & E     & What is the total number of nodes in the network?                                                                                                                                                                                                                                                                  \\
T2  & E     & Which devices in the network have the most IP addresses configured?                                                                                                                                                                                                                                                \\
T3  & E     & Which devices in the network have multiple IP addresses configured?                                                                                                                                                                                                                                                \\
T4  & E     & Which IPv6 addresses are configured in the network, and on which devices?                                                                                                                                                                                                                                           \\
T5  & M     & Which application services are running on each host? [...]                                                                                                        \\
T6  & M     & Are devices X1 and X2 directly connected? If so, on which subnet?                                                                                                                                                                                                                          \\
T7  & M     & Can machine X1 ping machine X2 directly without intermediate hops?                                                                                                                                                                                                                                \\
T8  & H     & Is zone transfer allowed on any DNS server? Which zones and to which hosts?                                                                                                                                                                                                                                          \\
T9  & H    & List all unique subnets in the network (in CIDR notation) [...] 
\\
T10 & H       & What is the traceroute from X1 to X2?         \\                                                                                                                           \bottomrule                                                                                                                      
\end{tabular}%
}%
\end{table}

The benchmark consists of 10 questions with progressively increasing difficulty, as summarized in Table~\ref{tab:questions}. The difficulty levels are categorized as \emph{Easy}, \emph{Medium}, and \emph{Hard}, reflecting the level of reasoning, abstraction, and interaction with the network required to produce a correct answer.

Easy questions (T1--T4) focus on basic inspection and retrieval operations over the network state. These tasks require identifying simple properties such as the total number of nodes (T1), detecting devices with multiple IP configurations (T2, T3), or listing IPv6 addresses and their associated devices (T4). Such questions primarily test the model's ability to extract and organize information without requiring complex reasoning or multi-step analysis.

Medium questions (T5--T7) introduce a higher level of complexity by requiring relational reasoning and partial interpretation of network behavior. For instance, identifying application services and special-purpose IP assignments (T5) involves correlating multiple pieces of information across hosts. Similarly, determining whether two devices are directly connected (T6) or whether direct communication is possible without intermediate hops (T7) requires an understanding of network topology and connectivity constraints.

Hard questions (T8--T10) are designed to assess advanced reasoning capabilities and the ability to interact with dynamic aspects of the network. These include security-related analyses, such as verifying DNS zone transfer configurations (T8), computing a minimal, non-redundant set of subnets (T9), and performing path-discovery operations, such as traceroute between two hosts (T10). These tasks typically require multi-step reasoning, interpretation of protocol behavior, and, in some cases, active probing of the network environment.


\subsection{Labs}

To evaluate the proposed methodology across heterogeneous scenarios, we consider six network labs with increasing levels of complexity. Each lab represents a distinct network architecture, characterized by different topologies, routing protocols, and deployed services. A summary is reported in Table~\ref{tab:kathara-labs}.

\begin{table*}[tbh]
\centering
\scriptsize
\caption{Overview of the Kathara-Labs used in the evaluation pipeline.}
\label{tab:kathara-labs}
\begin{tabular}{llllccc}
\toprule
\textbf{ID} & \textbf{Lab Name} & \textbf{Complexity} & \textbf{Description} & \textbf{\# Nodes} & \textbf{\# Subnets} \\
\midrule
%
L1 & \texttt{medium-ftp-dhcp-web} & Easy & Corporate networks with various services.   & 10 & 3 \\
L2 & \texttt{medium-nat-web} & Easy & Simple webserver accessible using NAT. & 11 & 4 \\
L3 & \texttt{dns-load-balancer-with-rip} & Medium & Network scenario with a load-balancer and RIP & 9 & 4 \\
L4 & \texttt{stairs} & Medium & Internet level networking with RIP and 4 ASs & 8 & 10 \\
L5 & \texttt{alien} & Difficult & Internet level networking with RIPv2 and 5 ASs & 10 & 9 \\
L6 & \texttt{small-internet-with-dns-and-web-server} & Difficult & Small Internet with DNS and a web server & 15 & 17 \\
\bottomrule
\end{tabular}
\end{table*}

The selected labs span from relatively simple enterprise-like environments to more complex, Internet-scale scenarios. The Easy labs (L1, L2) model typical corporate networks and basic service deployments. For instance, L1 includes a combination of common services such as FTP, DHCP, and web servers within a limited topology, while L2 introduces Network Address Translation (NAT) to expose internal services, adding an additional layer of abstraction.

The Medium labs (L3, L4) increase the complexity by incorporating routing protocols and more articulated topologies. L3 features a load-balanced service architecture combined with RIP-based routing, requiring a deeper understanding of traffic distribution and routing behavior. L4 further extends this setting by modeling an Internet-like environment with multiple autonomous systems (ASs) interconnected via RIP, significantly increasing the number of subnets and the overall structural complexity.

Finally, the Difficult labs (L5, L6) represent advanced scenarios that closely resemble real-world Internet configurations. These environments include multiple autonomous systems, more sophisticated routing protocols (e.g., RIPv2), and a higher node and subnet density. In particular, L6 combines DNS infrastructure with web services in a multi-subnet Internet setting, requiring comprehensive reasoning across both application- and network-layer components.


\subsection{Implemented Solvers} \label{solvers}
We implement four distinct solvers spanning the spectrum from a single-shot baseline to a fully agentic, tool-driven pipeline. These architectures differ along three main axes: (i) \emph{context management} (monolithic context injection vs.\ selective, iterative retrieval), (ii) \emph{reasoning structure} (direct answering vs.\ explicit chain-of-thought processing), and (iii) \emph{environmental interaction} (static file analysis vs.\ active tool execution). Each solver is orchestrated via a LangGraph\footnote{\url{https://langchain-ai.github.io/langgraph/}} workflow and enforces a structured response by emitting its final output as a Pydantic\footnote{\url{https://docs.pydantic.dev/latest/}} model matching the question's target schema.
\paragraph{Bulk Solver}
This baseline workflow establishes an efficiency and simplicity reference point, extending the monolithic approach of prior literature. The solver packs all network configuration files and the target question into a single comprehensive prompt. The underlying LLM must demonstrate end-to-end network understanding in a single forward pass without iterative refinement. The execution pipeline consists of two sequential nodes:
\begin{itemize}
    \item \textbf{\texttt{llm\_solver}}: Evaluates the global lab context alongside the query to generate a free-form textual response.
    \item \textbf{\texttt{structured\_output}}: Ingests the conversation history and applies a specialized system prompt to extract and map the answer into the required Pydantic schema.
\end{itemize}

\paragraph{Bulk+ReAct Solver}
This architecture augments the monolithic baseline by introducing an explicit, multi-step Reason--Act paradigm~\cite{yao2023reactsynergizingreasoningacting} while keeping the available upfront context identical to the Bulk Processing Solver. By decoupling reasoning from action within a shared state, it isolates the structural impact of explicit chain-of-thought planning. The workflow comprises three sequential LLM evaluations:
\begin{itemize}
    \item \textbf{\texttt{reason}}: Generates a step-by-step analytical breakdown of the network topology and the query requirements based on the raw context.
    \item \textbf{\texttt{act}}: Leverages the reasoning history to formulate a definitive, grounded answer.
    \item \textbf{\texttt{structured\_output}}: Assembles the full conversation history to synthesize the final Pydantic model.
\end{itemize}

\begin{figure}[tb]
    \centering
    \subfloat[Guided Retrieval Agent Solver workflow]{%
        \includegraphics[height=6cm]{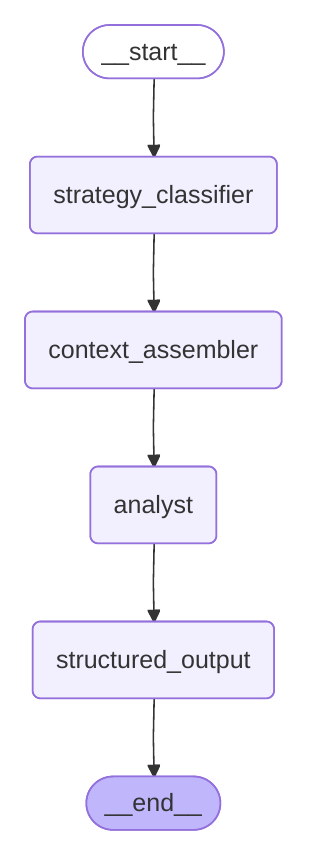}%
        \label{fig:guided_retrieval}}
    \hspace{1cm}
    \subfloat[Planner Agent Solver workflow]{%
        \includegraphics[height=6cm]{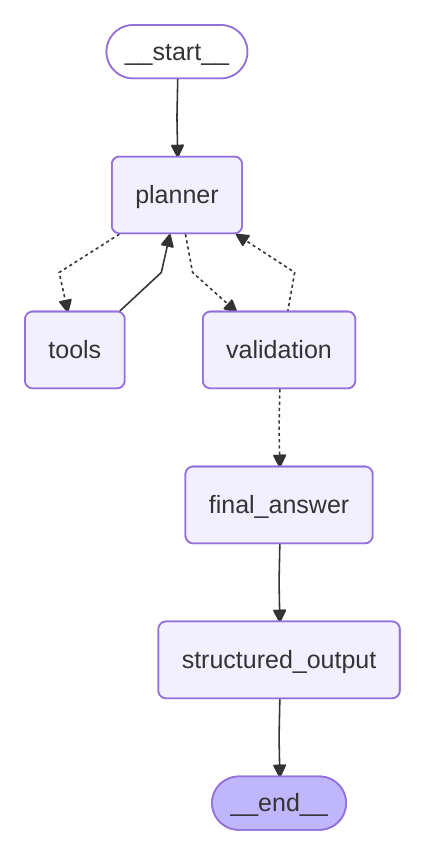}%
        \label{fig:planner_agent}}
    \caption{Architecture comparison between our agentic solvers. (a) The Guided Retrieval Agent decouples retrieval via a single discrete classification step. (b) The Planner Agent drives an iterative, tool-mediated retrieval loop governed by an automated validation gate.}
    \label{fig:agent_solvers}
\end{figure}

\paragraph{Guided Retrieval Agent Solver}
We propose this approach (Fig.~\ref{fig:agent_solvers}(a)) to mitigate the fragility and non-determinism of multi-turn tool-calling loops, which often cause small or local LLMs to hallucinate or emit empty responses. This lightweight agentic architecture decouples the high-level retrieval strategy from the core reasoning task by replacing open-ended tool searching with a deterministic execution pipeline:
\begin{itemize}
    \item \textbf{\texttt{strategy\_classifier}}: An initial LLM node maps the question onto one of five domain-specific retrieval strategies: \texttt{topology\_only}, \texttt{ip\_analysis}, \texttt{device\_pair}, \texttt{live\_connectivity}, or \texttt{service\_scan}.
    \item \textbf{\texttt{context\_assembler}}: A deterministic software layer that selectively parses only the subset of laboratory configuration files relevant to the chosen strategy, eliminating LLM tool-binding overhead.
    \item \textbf{\texttt{analyst} \& \texttt{structured\_output}}: A tool-free LLM node interprets the highly curated, strategy-specific context, passing its findings to the final node for schema extraction (with a fallback mechanism to the assembled context if the analyst's output is blank).
\end{itemize}
This setup bounds the operational cost to a fixed budget of three LLM calls while preserving strategic flexibility.

\paragraph{Planner Agent Solver}
This is the most sophisticated pipeline implemented: an autonomous, tool-using system governed by an explicit planner-validator feedback loop (Fig.~\ref{fig:agent_solvers}(b)). Instead of relying on predefined ingestion paths, the agent dynamically drives its own data-gathering process:
\begin{itemize}
    \item \textbf{\texttt{planner}}: Dynamically assesses the information gap and invokes localized tools to inspect the network environment, maintaining an execution loop until it flags the internal state as complete.
    \item \textbf{\texttt{tools}}: Executes specific programmatic hooks. In the file-only variant, the agent accesses utilities such as \texttt{list\_lab\_files} and \texttt{read\_device\_config}. 
    \item \textbf{\texttt{validation}}: Acts as an automated quality gate, verifying whether the accumulated context satisfies every field of the target output schema. If deficient, it routes execution back to the \texttt{planner} with feedback notes; execution is strictly bounded by a validation retry threshold and a total recursion limit to prevent infinite loops.
    \item \textbf{\texttt{final\_answer} \& \texttt{structured\_output}}: Upon validation success, these nodes synthesize the multi-turn evidence into a natural language response and extract the final Pydantic model, respectively.
\end{itemize}

\subsection{Foundation Models} \label{sec:foundation-models}
We evaluate two complementary tiers of foundation models, detailed in Table~\ref{tab:models-full}. The first consists of \emph{locally deployed open-weight models}, small enough to run on commodity or on-premises hardware; all are served through Ollama at 4-bit \texttt{Q4\_K\_M} quantization. The second consists of \emph{large-scale models accessed through APIs}, which we adopt as an upper-bound reference. For every model, we report whether it natively advertises \emph{tool/function calling} (\emph{Tools}) and an explicit \emph{reasoning} mode (\emph{Reason}); for the local tier, these flags are taken directly from the capabilities exposed by Ollama.

\begin{table}[tb]
    \centering
    \scriptsize
    \label{tab:models-full}
    \setlength{\tabcolsep}{3pt}
    \caption{Evaluated language models. Locally-deployed open-weight models are served via Ollama; large-scale models are accessed via provider APIs.}
    \label{tab:models}
    \begin{threeparttable}
    \resizebox{\columnwidth}{!}{%
    \begin{tabular}{@{}cllccccc@{}}
      \toprule
      & Model & Dev. & Params & Quant & Tools & Reason. & Year \\
      \midrule
      
      \multirow{8}{*}{\rotatebox[origin=c]{90}{Local}}
      & LFM2.5-Thinking~\cite{liquidai2025lfm2} & Liquid AI  & 1.2B                  & Q4\_K\_M & \checkmark & \checkmark & 2026 \\
      & Granite 4~\cite{granite2025}       & IBM        & 3.4B                  & Q4\_K\_M & \checkmark & --         & 2025 \\
      & Nemotron 3 Nano~\cite{blakeman2025nvidia} & NVIDIA     & 4.0B                  & Q4\_K\_M & \checkmark & \checkmark & 2025 \\
      & Llama 3.1~\cite{grattafiori2024llama}       & Meta       & 8.0B                  & Q4\_K\_M & \checkmark & --         & 2024 \\
      & Gemma 4 E4B~\cite{google2026gemma4}     & Google     & 8.0B\tnote{$\dagger$} & Q4\_K\_M & \checkmark & \checkmark & 2026 \\
      & RNJ-1~\cite{rnj1_5_instruct}           & Essential AI & 8.3B                & Q4\_K\_M & \checkmark & --         & 2025 \\
      & Qwen 3.5~\cite{qwen3}        & Alibaba    & 9.7B                  & Q4\_K\_M & \checkmark & \checkmark & 2026 \\
      & Ministral 3~\cite{liu2026ministral}     & Mistral AI & 13.9B                 & Q4\_K\_M & \checkmark & --         & 2025 \\

      \midrule

      \multirow{2}{*}{\rotatebox[origin=c]{90}{API}}
      & Kimi K2.5~\cite{kimiteam2026kimik25visualagentic}       & Moonshot AI & 1T / 32B   & native & \checkmark & \checkmark & 2026 \\
      & GLM-5~\cite{glm5team2026glm5vibecodingagentic}           & Z.ai        & 744B / 40B & native & \checkmark & \checkmark & 2026 \\

      \bottomrule
    \end{tabular}%
    }%
    \begin{tablenotes}[flushleft]
      \item[$\dagger$] 8.0B total / $\sim$4B effective parameters (elastic architecture).
    \end{tablenotes}
    \end{threeparttable}
\end{table}

Whereas the local tier spans roughly $1$-$14$B parameters, the API tier operates in the hundreds-of-billions to trillion-parameter regime. 
This contrast allows us to quantify the performance gap between on-premises deployable models and frontier systems on our network topology reasoning tasks.

\subsection{Evaluation Metrics}
To systematically assess the performance of each setup, we defined a set of quantitative evaluation metrics and evaluated each experiment along three axes: \emph{correctness}, \emph{efficiency}, and \emph{trajectory behavior}.

\textbf{Correctness}: The proportion of correctly answered questions out of the total attempted for each model/solver pair. 
To compute that, each run is scored as a binary outcome based on whether the solver's structured output matches the ground-truth state of the network emulation environment. Each experimental run is classified into one of three mutually exclusive outcomes: (i) \emph{Correct}, if the output matches the ground truth; (ii) \emph{Wrong}, if the execution completes and returns a syntactically valid schema, but with incorrect semantic values; and (iii) \emph{Invalid Run}, if the pipeline fails to deliver a format-compliant answer. Invalid runs are further categorized based on the specific failure trigger, such as \emph{Empty Response} (the model fails to emit a final answer string), \emph{Malformed JSON} (severe syntax parsing violations), or \emph{Schema Violation} (valid JSON that violates the expected Pydantic data model constraints).
    
\textbf{Efficiency:} 
The framework monitors both the computational and temporal resource footprint of the system. Operational costs and optimization are assessed by calculating the \emph{Average Token Usage} per task and analyzing how token consumption scales across different models and solvers. This is paired with the \emph{Average Execution Latency}, which computes the mean wall-clock time required by each model-solver configuration to process a query and generate its final structured output.

\textbf{Trajectory}: To understand the inner workings of agentic exploration and map exactly \emph{how} solvers navigate the network environment, the framework captures the detailed execution path of each run through fine-grained trajectory metrics:
\begin{itemize}
        \item \emph{Failure Stage}: Identifies the exact node within the LangGraph workflow where a failure or an exception is triggered 
        allowing to pinpoint the root cause independently of the terminal schema extraction.
        \item \emph{Tool Interaction}: For tool-using solvers, the number of tool invocations, failed calls, and repeated identical calls per run, characterizing whether retrieval is productive or degenerates into thrashing.
        \item \emph{Exploration Effort}: The number of planner iterations and of distinct configuration files read before answering, quantifying how much work a configuration expends to reach its result.
        \item \emph{Wrong Answer analysis}: For \emph{Wrong Answer} runs only, the structural class of the discrepancy with the ground truth, distinguishing value-level errors (correct structure, wrong values) from structural ones (missing, extra, or jointly altered items) and from domain-specific rejections (e.g., an invalid traceroute path).
    \end{itemize}

These metrics collectively provide a comprehensive framework for evaluating both the correctness and efficiency of LLM-based approaches to network administration. They enable detailed comparison across models, solvers, laboratory scenarios, and question types.

\subsection{Experimental settings}
We adopt a full-factorial design over the 4 solvers
(Section~\ref{solvers}), the 10 foundation models of
Section~\ref{sec:foundation-models}, the 10 questions
(Table~\ref{tab:questions}), and the 6 labs
(Table~\ref{tab:kathara-labs}). Each $(\texttt{solver},\texttt{model},\texttt{question},\texttt{lab})$ configuration is
executed ten times, for a total of $24{,}000$ runs. For device-specific
questions, the target devices are fixed in advance so as to guarantee a
non-trivial ground truth (e.g., device pairs that are not adjacent on an obvious
shared subnet).

The 8 open-weight models are served locally through
Ollama\footnote{\url{https://ollama.com/}} at 4-bit \texttt{Q4\_K\_M}
quantization on a single commodity workstation (
Intel
Core i9-12900F, 24 threads; 32\,GB RAM; NVIDIA GeForce RTX~4070~Ti, 12\,GB
VRAM), reflecting a realistic on-premise deployment budget. The two large-scale
models, Kimi~K2.5 and GLM-5, are accessed through Amazon Bedrock. Both tiers are
driven by the same LangGraph orchestration layer, and every solver emits its
final answer as a Pydantic model. Network environments are instantiated with the
Kathará emulator~\cite{bonofiglio2018kathara}. The Planner Agent solver is
bounded by a validation-retry threshold of $3$ and a recursion limit of $25$ to
prevent non-terminating loops.

\section{Results}
\subsection{Overall Performance}\label{sec:rq1}


\begin{figure}[tb]
    \centering
    \includegraphics[trim={15 23 15 25},clip,width=\linewidth]{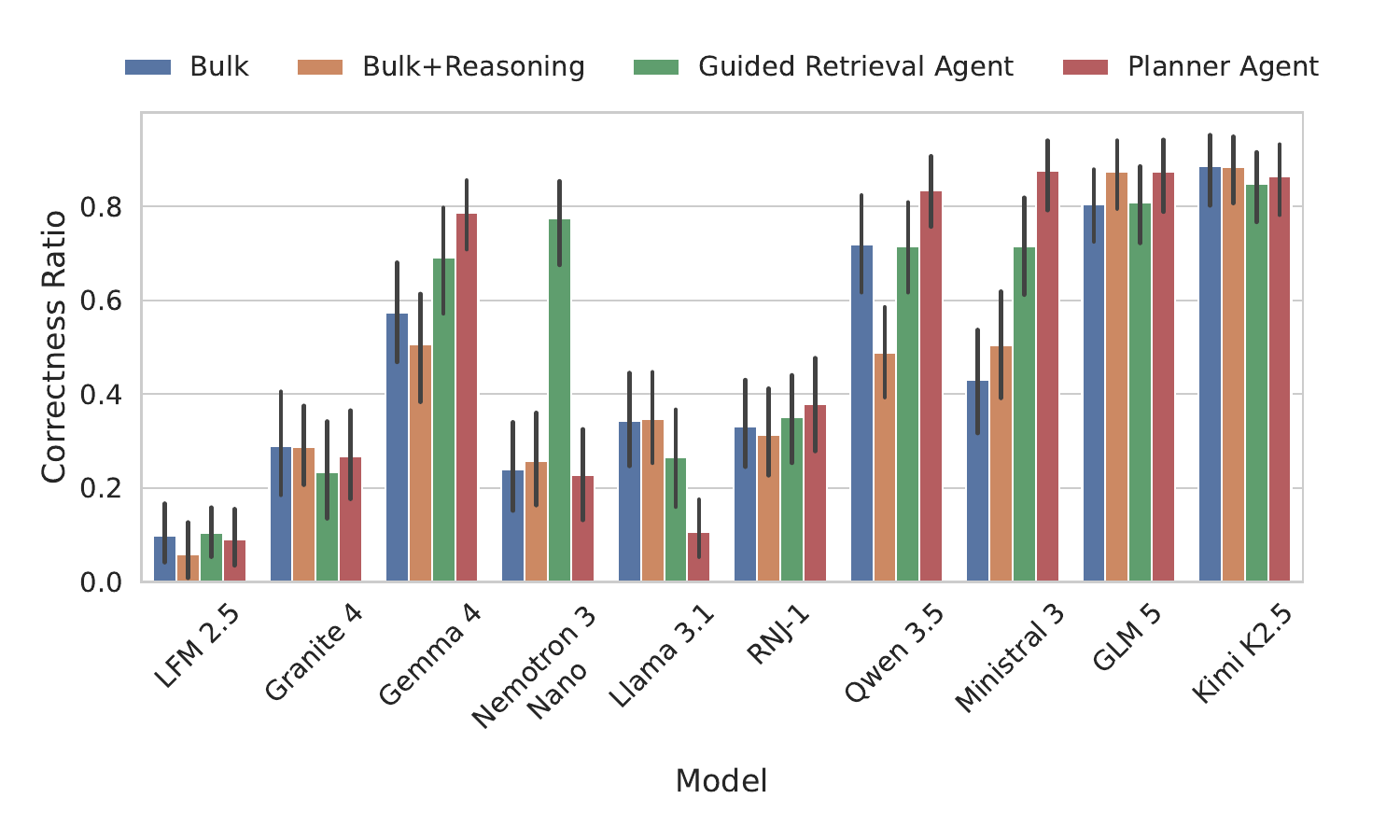}
    \caption{Correctness ratio of every model--solver pair, with models ordered
    by parameter count from LFM~2.5~Thinking (1.2B) to the trillion-parameter
    Kimi~K2.5 reference. Error bars denote confidence intervals across labs and
    questions.}
    \label{fig:overall}
\end{figure}

Fig.~\ref{fig:overall} reports the correctness ratio of every model--solver
pair, with models ordered by ascending parameter count. 
The headline result is that competitive network-reasoning performance does not require frontier-scale models: the best local configuration is statistically indistinguishable from the upper-bound anchor while remaining small enough to run on commodity hardware. Under the Planner Agent solver, Ministral~3 (14B) reaches a correctness ratio of $0.88$, matching Kimi~K2.5's best result ($0.88$), and Qwen~3.5 (9B) follows closely at
$0.83$; in all three cases, the differences fall well within the confidence
intervals. Since local models are served at 4-bit quantization, this parity is a favorable trade for the privacy, cost, and latency benefits
of on-premise deployment.
 
Performance does scale with model size, but only loosely, and the relationship is markedly non-monotonic. At the low end, LFM~2.5~Thinking (1.2B) is effectively non-functional across every solver (correctness ratios below $0.10$), confirming that a minimum capacity threshold must be met before any orchestration strategy becomes useful. Above that threshold, however, architecture and training quality dominate raw parameter count: Gemma~4~E4B, with only $\sim$4B effective parameters, consistently outperforms the larger Llama~3.1 (8B) and RNJ-1 (8B). 
The eight local models, therefore, do not form a clean size-ordered ranking, indicating that, in this regime, \emph{which} model is chosen matters more than \emph{how large} it is.
 
Finally, GLM~5 and Kimi~K2.5 are distinguished less by their peak accuracy than by their robustness: they are the only models to remain strong across all four solvers (correctness ratios of $0.80$--$0.88$), whereas the local models are far more sensitive to the choice of orchestration strategy. 
This sensitivity
motivates the per-solver analysis in Section~\ref{sec:rq2}, where we examine how orchestration interacts with model capability.

\subsection{Effect of Solver Architecture}\label{sec:rq2}
 
\begin{figure}[tb]
    \centering
    \includegraphics[width=.8\linewidth]{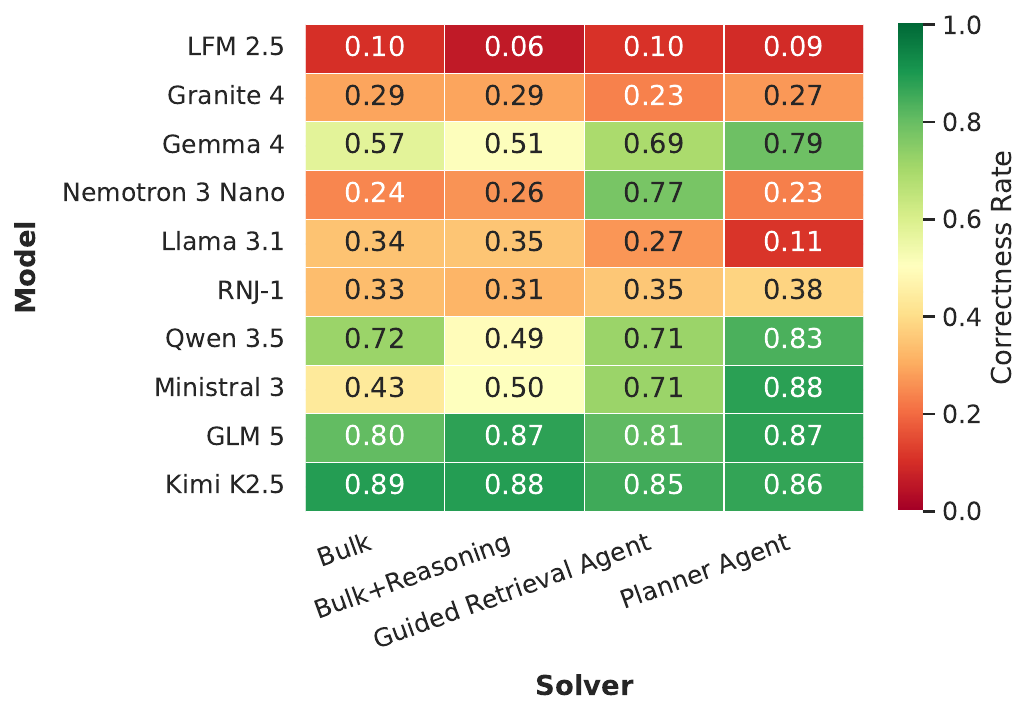}
    \caption{Correctness rate of each model--solver pair, aggregated over all labs and questions. Models are ordered by parameter count. 
    }
    \label{fig:rq2_heatmap}
\end{figure}
 
Fig.~\ref{fig:rq2_heatmap} reports the correctness rate of every model--solver pair, aggregated across all labs and questions. The effect of a given solver is not uniform across models: the Planner Agent yields the highest scores for the larger, more capable models, raising Ministral~3 (14B) from $0.43$ under the Bulk baseline to $0.88$ and also improving Qwen~3.5 (9B, $0.72 \rightarrow 0.83$) and Gemma~4~E4B ($0.57 \rightarrow 0.79$); for smaller models, however, the same tool-driven loop can be detrimental, as with Llama~3.1 (8B), which drops from $0.34$ to $0.11$.
 
The Guided Retrieval Agent exhibits a complementary profile. Its most pronounced effect is on Nemotron~3~Nano (4B), from $0.24$ to $0.77$ (the single
largest improvement of any pair), and it also lifts Ministral~3 from $0.43$ to $0.71$, while leaving other models neutral or slightly worse. In contrast, the
Bulk+ReAct solver, which adds an explicit reasoning step over the same upfront context, produces little improvement and in several cases a reduction in
correctness (e.g., Qwen~3.5 from $0.72$ to $0.49$, Gemma~4~E4B from $0.57$ to $0.51$).
 
Finally, GLM 5 and Kimi~K2.5 remain close to constant across all four solvers ($0.80$--$0.89$), showing little sensitivity to orchestration. Overall, the benefit of a more elaborate solver depends strongly on the underlying model: agentic pipelines provide the largest gains for capable models, while offering limited or negative returns for the weakest models and for models already strong under the simplest baseline.

\subsection{Performance Across Task Types}\label{sec:rq3}
 
\begin{figure}[tb]
    \centering
    \includegraphics[width=\linewidth]{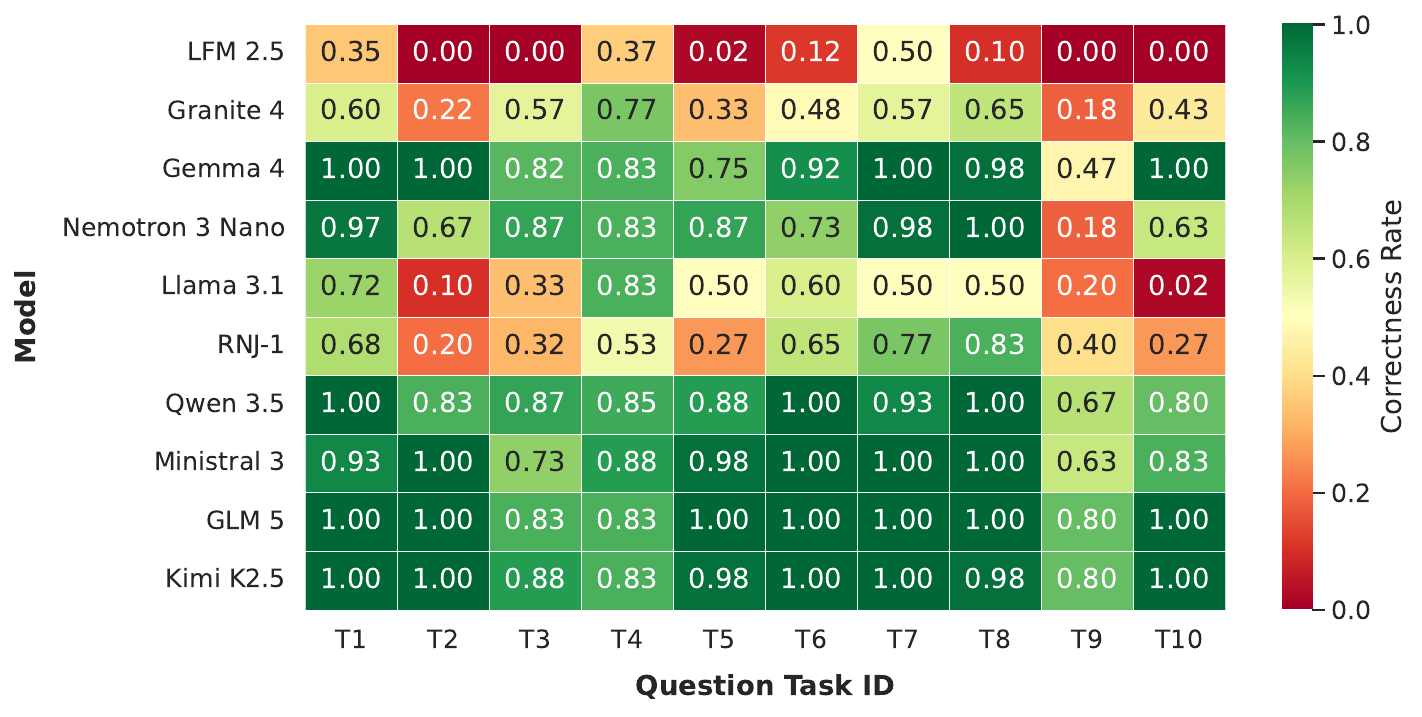}
    \caption{Correctness rate of each model on each question task, using the best solver per model and averaging over the labs where the task applies.
    Models are ordered by size; tasks T1--T10 by their a priori complexity label.}
    \label{fig:rq_model_question}
\end{figure}
 
Fig.~\ref{fig:rq_model_question} shows that empirical task difficulty does not always match the a priori Easy/Medium/Hard labeling. At the hard end, T9 
(computing a minimal, non-redundant set of subnets) 
is the clear bottleneck for every model---even Kimi~K2.5 ($0.80$) and Ministral~3 ($0.63$) decline, while most local models fall below $0.20$---whereas the other two Hard tasks are solved routinely (T8, DNS zone transfer, reaches $0.98$--$1.00$ for capable models, and T10, traceroute, is high across the stronger tier). 
A symmetric mismatch appears among the Easy tasks: the comparative tasks T2 and T3 
trip up weaker models such as Llama~3.1 ($0.10$, $0.33$) and RNJ-1 ($0.20$, $0.32$), despite simple retrieval tasks like node counting (T1) being solved broadly. 
By row, the capability frontier is as expected, with Qwen~3.5, Ministral~3, and large models like GLM 5 and Kimi~K2.5 near-perfect everywhere. Overall, difficulty is driven less by the coarse complexity category than by the specific reasoning operation involved, with set minimization (T9) and comparative selection (T2/T3) being disproportionately hard relative to their labels.

\subsection{Best Deployable Configuration Across Scenarios}\label{sec:rq-casestudy}
 
\begin{figure}[tb]
    \centering
    \includegraphics[trim={0 0 0 48},clip,width=\linewidth]{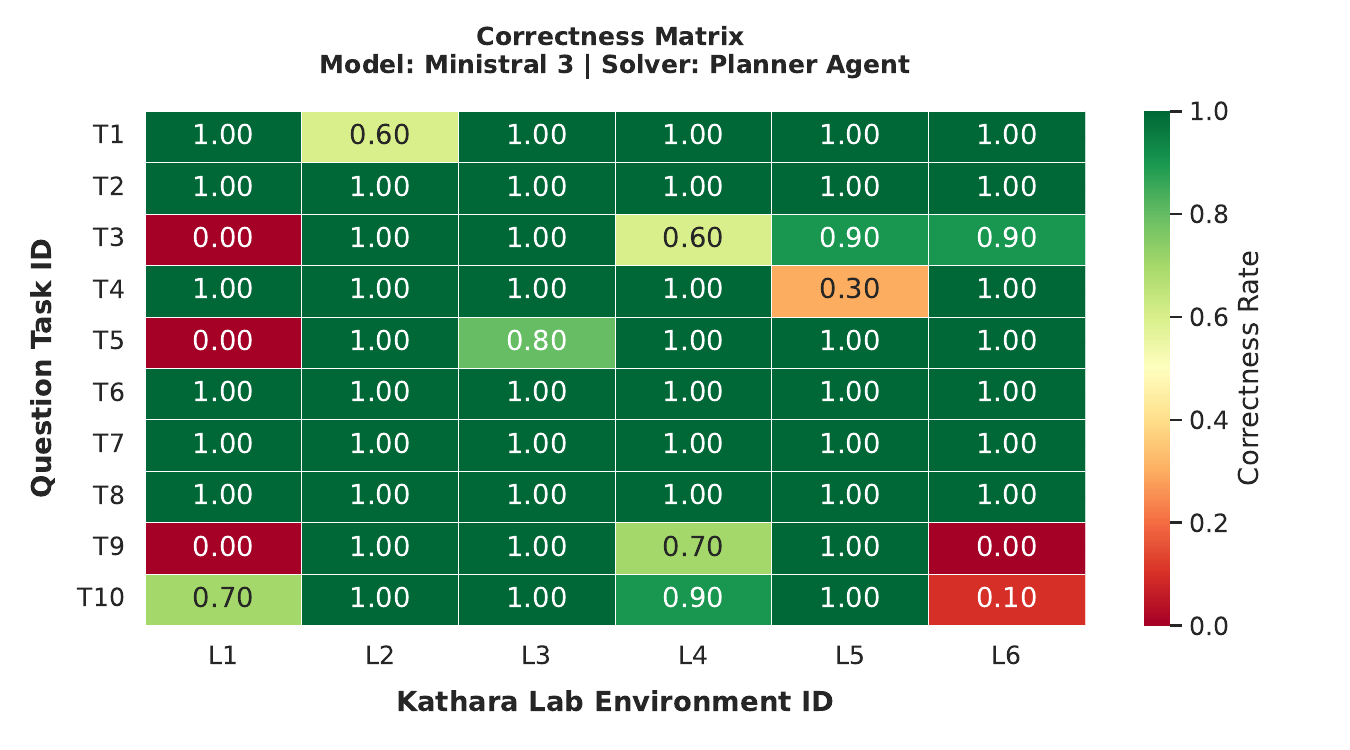}
    \caption{Correctness matrix for the best deployable configuration (Ministral~3~14B with the Planner Agent solver), broken down by question task (T1--T10) and lab environment (L1--L6). Each cell is averaged over the repetitions of the corresponding (task, lab) pair; blank cells denote tasks that do not apply to a given lab.
    }
    \label{fig:rq_task_lab}
\end{figure}
 
To examine how the strongest locally deployable configuration behaves across network scenarios of increasing complexity, Fig.~\ref{fig:rq_task_lab} reports the correctness matrix for Ministral~3~(14B) under the Planner Agent solver, resolved by task and by lab. The configuration is highly effective overall, with the large majority of cells fully solved across all six labs, including the Internet-scale scenarios L5 and L6. Notably, performance does not degrade monotonically with network complexity: tasks such as connectivity and reachability reasoning (T6, T7) and DNS zone-transfer analysis (T8) are solved perfectly even on the largest topologies, indicating that for this configuration scenario size alone is not the dominant difficulty factor.
 
Instead, the residual failures are concentrated in a small number of specific (task, lab) combinations rather than spread along a complexity gradient. The minimal-subnet task T9 is solved on the mid-complexity labs (L2, L3, L5) yet fails on both the simplest lab L1 ($0.00$) and the most complex lab L6 ($0.00$), a non-monotonic pattern that a coarse complexity ordering does not explain.
Path discovery (T10) is the other notable case, collapsing specifically on the 17-subnet, multi-AS lab L6 ($0.10$) while remaining reliable elsewhere. A few further isolated dips (e.g.\ T3 and T5 on L1) follow the same localized character. Because these anomalies are tied to particular task--scenario pairs rather than to the overall difficulty, we defer their explanation to a dedicated per-trajectory failure analysis (Section~\ref{sec:trajectory-analysis}), which inspects the recorded execution traces and structured output differences to
determine whether each case reflects a model reasoning failure or an artifact of ground-truth comparison.

\subsection{Token and Latency Cost}\label{sec:rq-cost}
 
Beyond correctness, the four solvers differ substantially in their resource footprint. Averaged across all models, the Guided Retrieval Agent is the most token-efficient solver ($\sim$5{,}700 tokens per task), followed by Bulk+ReAct ($\sim$7{,}100) and the Bulk baseline ($\sim$9{,}200); both structured solvers are cheaper than the monolithic baseline, as they either curate or compress the context rather than injecting every configuration file. The Planner Agent occupies a separate cost regime, consuming $\sim$30{,}800 tokens per task on average, roughly $3.3\times$ the Bulk baseline and over $5\times$ the Guided Retrieval Agent. Latency follows a similar trend: Bulk and Guided Retrieval are the fastest ($\sim$16\,s on average), Bulk+ReAct is intermediate ($\sim$27\,s), and the Planner Agent is the slowest ($\sim$40\,s), with peak per-model latencies exceeding $75$\,s.
 
The Planner Agent's overhead is also strongly model-dependent: its token multiplier relative to the Bulk baseline ranges from $0.7\times$ for Llama~3.1, which exits the planning loop early, to $5.4\times$ for Granite~4 and Nemotron~3 Nano, which consume the most tokens of any configuration ($\sim$44{,}000 and $\sim$53{,}000 respectively). 
Models that already perform well with simpler solvers inflate far less. Kimi~K2.5, for instance, grows only $1.7\times$ under the Planner. The additional cost of the agentic pipeline is borne disproportionately by the weaker models that benefit from it the least. 
Last, local and API models exhibit comparable token counts and latencies across solvers, indicating that in our setup, solver choice, rather than deployment tier, is the dominant factor in overall cost. 

\subsection{Trajectory Analysis}\label{sec:trajectory-analysis}
Aggregate correctness tells us how often a configuration succeeds; the recorded execution traces tell us how and why it fails. Of the $24{,}000$ runs, $50.1\%$ are \emph{Correct}, $42.9\%$ are \emph{Wrong}---schema-valid outputs whose values do not match the ground truth---and $7.6\%$ are \emph{Invalid Runs}. The incorrect share varies enormously with capability---$85.7\%$ of runs for the smallest model, LFM~2.5~Thinking, but only $10.2\%$ for Kimi~K2.5 and $16.0\%$ for GLM-5---whereas invalid runs concentrate in a handful of models, dominated by Llama~3.1 (over $40\%$ of its runs), RNJ-1 ($\sim$$15\%$), and Qwen~3.5 ($\sim$$9\%$).

Invalid runs fall into two categories. The most impactful is silence (an \emph{empty response}): rather than answering incorrectly, the weakest models simply fail to produce an answer. Llama~3.1 returns nothing on nearly two runs in five, and RNJ-1 on roughly one in ten. These empty replies are produced while the model is still reasoning or drafting, not lost at the final formatting step: over half occur at the Planner's answer-synthesis stage, and a further $40\%$ at the analyst node (Fig.~\ref{fig:guided_retrieval}) of the Guided Retrieval solver. Under the Planner, roughly three-quarters of these cases abandon the task before issuing even a single tool call. Llama~3.1 is the extreme: it never invokes a tool under the Planner and instead piles up empty turns, which is precisely why the tool-driven pipeline lowers its accuracy ($0.34 \rightarrow 0.11$). The second category, a formatting or extraction fault, is markedly solver-dependent rather than intrinsic: Qwen~3.5's malformed outputs climb to roughly one run in eight under the reasoning-augmented Bulk solver yet fall below $2\%$ under the Planner.

Of the $7.8\%$ of runs that never yield a usable answer, about $70\%$ break at the answer-production stage and a further $25\%$ are faults in the structured output itself (broken JSON or schema violations), surfacing at the final extraction node common to every solver---confirming that \emph{where} a run breaks is dictated by the solver, while \emph{whether} its content is right is a property of the model. Restricting to the Planner, the only tool-driven solver, models that engage the loop issue on average 6--10 tool calls per run, and for six of the eight that engage at all, incorrect runs issue \emph{more} calls than correct ones ($1.2\times$ to over $2\times$)---a signature of search that fails to converge. The two weakest models behave oppositely, issuing almost no Planner tool calls per run---$0.5$ for LFM~2.5~Thinking and $0.0$ for Llama~3.1---and answering largely without evidence. Erroneous tool interactions are otherwise rare and localized: on \emph{Wrong} Planner runs, only RNJ-1 exhibits them at scale, averaging $9.5$ failed tool calls and $12.0$ repeated identical tool calls per run---an order of magnitude above every other model.

More revealing than the failures is how comparable accuracy can be achieved with different amounts of work. Ministral~3, GLM-5, and Kimi~K2.5 finish within a point of one another ($87.5\%$, $87.3\%$, $86.3\%$) yet diverge in evidence gathering: GLM-5 and Kimi~K2.5 issue nearly identical tool-call and iteration budgets (averaging about $9.5$ tool calls and $3.4$ planner iterations per run), but Kimi~K2.5 reads more than twice as many configuration files ($5.9$ vs.\ $2.7$), and Ministral~3 reads more still ($6.4$) over more iterations. Granite~4 shows the opposite anti-pattern, averaging $6.0$ tool calls over $7.0$ iterations per run, yet reads only $0.1$ configuration files per run and remains wrong on $73\%$ of runs. Finally, incorrect runs consistently call more tools and read more files than correct ones (e.g., GLM-5 $2.7 \rightarrow 4.6$, Kimi~K2.5 $5.9 \rightarrow 8.7$), confirming that extra exploration triggered by a hard instance does not translate into a correct answer.









\section{Discussion}\label{sec:discussion}

Below a certain capacity threshold, no orchestration strategy compensates for model weakness~\cite{donadel2024can,liu2024llmnetworkingworkflow}. Above it, architecture and training quality predict performance better than parameter count: \emph{which} model is chosen matters more than \emph{how large} it is. Task difficulty also proves an unreliable guide: empirical hardness is driven by the specific reasoning operation required, with set minimization proving disproportionately hard and DNS analysis disproportionately easy relative to their labels. Topology complexity is a similarly weak predictor in isolation: for the strongest deployable configuration, failures cluster in specific task--lab combinations rather than scaling with network size, with some hard tasks solved perfectly even on the largest topologies. Benchmark designers should therefore characterize tasks by reasoning operation rather than topology scale.
\begin{formal}
\textbf{RQ1---Baseline Capability.} 
Solver architecture matters more than LLM parameter count; difficulty depends mainly on the reasoning operation, not topology scale.
\end{formal}

Solver choice and model capability interact rather than add: a more sophisticated solver acts as a multiplier, amplifying the strengths of capable models and the weaknesses of weaker ones. The Planner Agent hurts models that lack the capacity to drive a tool loop, which instead produce empty intermediate turns~\cite{huang2024understanding}. The Guided Retrieval Agent avoids this by using a deterministic retrieval pipeline, making it the safer default for mid-range models. Bulk+ReAct~\cite{yao2023reactsynergizingreasoningacting} rarely improves over the monolithic baseline, suggesting that the ReAct paradigm's gains in other domains rely on curated rather than bulk-injected context. Frontier models remain robust across all four solvers, confirming that sufficient LLM size makes the choice of orchestrator largely irrelevant.
\begin{formal}
\textbf{RQ2---Effect of Solver Architecture.} Solver sophistication is a capability multiplier, not an additive boost; matching solver architecture to model capability is as consequential as the solver design itself.
\end{formal}

Under the right configuration, on-premise open-weight models close the accuracy gap with frontier API services to within confidence intervals, addressing the privacy and cost barriers that limit adoption of LLM-based network management~\cite{hong2025comprehensive,wu2024netllm,donadel2024can}. The parity is conditional, however: local models are far more sensitive to solver choice than frontier models, so the effort saved on API costs must be reinvested in solver design. Notably, this reinvestment does not come for free at the resource level either: local and API-served models incur comparable token and latency costs once a given solver is fixed, since solver choice, rather than deployment tier, is the dominant driver of resource consumption. The benefit of going local is therefore data privacy and the elimination of per-token API fees, not a reduction in computational effort. Competitive on-premise accuracy is achievable, but only when solver selection is treated as a deliberate engineering decision.
\begin{formal}
\textbf{RQ3---Local versus Frontier Models.} Locally-deployable open-weight models can match frontier-scale API models in accuracy under the right solver and eliminate per-token API costs and privacy exposure, though resource consumption itself is governed by solver choice rather than deployment tier. 
\end{formal}

The Guided Retrieval Agent is both the most token-efficient solver and the strongest performer for mid-range models, dominating the cost--accuracy frontier. Bulk+ReAct is the weakest trade-off: it costs less than the Bulk baseline yet delivers inconsistent accuracy returns. The Planner Agent's overhead ($3.3\times$ Bulk on average) falls disproportionately on the weaker models that benefit from it least. More broadly, efficiency and correctness are largely orthogonal: using more tokens does not mean a model is reasoning more effectively; in many cases, it simply means the model is stuck, exploring without getting any closer to the answer.
\begin{formal}
\textbf{RQ4---Cost--Accuracy Trade-offs.} The Guided Retrieval Agent dominates on the cost--accuracy frontier; the Planner Agent's overhead is worthwhile only for capable models. 
\end{formal}

Three distinct failure patterns emerge from the trajectory analysis.
The most impactful is \emph{silence}: the dominant failure mode for weak models is producing no answer rather than a wrong one, meaning aggregate accuracy metrics understate their true unreliability. The second is \emph{thrashing}: for the majority of models that engage the Planner's tool loop, incorrect runs issue more tool calls than correct ones (from $1.2\times$ to over $2\times$ more), with calls cycling without converging on new evidence.
The third is \emph{blind looping}: a model that actively engages the planning loop yet reads almost no configuration files, answering largely without evidence---illustrated by Granite~4, which averages six tool calls over seven iterations while reading fewer than one file per run and finishing $73\%$ incorrect. Formatting faults, by contrast, are solver-dependent rather than model-intrinsic, identifying structured-output extraction as a fragile integration point that solver design should explicitly harden~\cite{huang2024understanding}. Orchestration shapes how a model fails, but cannot supply the understanding it lacks.
\begin{formal}
\textbf{RQ5---Failure Characterization.} Solver architecture determines \emph{how} and \emph{where} a run fails; model capability determines whether it is correct. Weak models fail through silence, agents through thrashing or blind looping.
\end{formal}

\section{Conclusions}

We presented \name{}, a framework for automatically evaluating LLM-based pipelines on network administration tasks. By grounding every evaluation in a live emulated network, \name{} removes the need for manual validation and generates programmatic, scenario-specific ground truth at scale. A full-factorial study of $24{,}000$ runs across 10 foundation models, 4 solver architectures, 10 task types, and 6 network labs of increasing complexity yields five concrete findings: a sharp capacity floor below which no solver compensates for model weakness; architecture and training quality as stronger predictors than parameter count above it; solver sophistication as a capability multiplier rather than an additive boost; parity between local and frontier models under the right configuration; and three distinct trajectory failure patterns that aggregate accuracy alone cannot distinguish.

Taken together, these results suggest that the primary bottleneck in LLM-based network management is no longer model scale but solver--model alignment: selecting and designing the orchestration strategy to match the capacity of the underlying model yields gains that additional parameters alone do not. The open-source release of \name{} is intended to make this line of investigation reproducible and extensible.

Several directions remain open. The current benchmark covers read-only reasoning tasks; extending it to include configuration generation, access-control auditing, and fault remediation would significantly broaden coverage. The live-network probing capability of \name{} is implemented but was held constant across experiments; future work should study how active probing interacts with solver design and task type. Finally, the robustness gap between local and frontier models across solver choices motivates exploring lightweight fine-tuning of open-weight models on networking corpora as a complement to solver engineering.


%




\ifCLASSOPTIONcaptionsoff
  \newpage
\fi



\bibliographystyle{IEEEtran}
\bibliography{bib}
%

%




\begin{IEEEbiographynophoto}{Gianmaria Frigo} is currently pursuing the M.Sc. degree in Computer Engineering at the University of Padua, Italy, with a specialization in artificial intelligence and robotics. He received the B.Sc. degree in Computer Engineering from the University of Padua in 2025. His interests include computer vision, deep learning, multimodal perception, and autonomous robotic systems.
\end{IEEEbiographynophoto}

\vspace{-3em}

\begin{IEEEbiographynophoto}{Davide Saladino} holds a Bachelor's degree in Computer Engineering from the University of Padua, where he is currently pursuing a Master's degree in Computer Science. He works as a cybersecurity professional and is deeply passionate about the field.
\end{IEEEbiographynophoto}

\vspace{-3em}

\begin{IEEEbiographynophoto}{Alberto Castagnaro} received the MSc degree in Computer Science at TU Delft, The Netherlands, in 2024. He is a cybersecurity Engineer at Spritz Matter. His research covers LLM security and AI-enhanced red-teaming tools.  
\end{IEEEbiographynophoto}

\vspace{-3em}

\begin{IEEEbiographynophoto}{Francesco Marchiori} received the Ph.D. degree in Brain, Mind, and Computer Science from the University of Padua, Italy, in 2025. His research interests include artificial intelligence applied to cybersecurity, security and privacy of cyber-physical systems, adversarial machine learning, cyber threat intelligence, and automotive security.
\end{IEEEbiographynophoto}

\vspace{-3em}

\begin{IEEEbiographynophoto}{Denis Donadel} received the Ph.D. from the University of Padua in 2024. He is a postdoctoral researcher at the University of Verona, Italy. His research interests include cybersecurity of cyber-physical systems, automotive security, and AI applied to cybersecurity. 
\end{IEEEbiographynophoto}

\vspace{-3em}

\begin{IEEEbiographynophoto}{Luca Pajola} received is PhD at the University of Padova in 2023. He is a Secure AI Engineer at Spritz Matter SRL, a spin-off of the University of Padova. His research interests include AI security, network intrusion detection, and adversarial machine learning, 
\end{IEEEbiographynophoto}

\vspace{-3em}

\begin{IEEEbiographynophoto}{Mauro Conti} (Fellow, IEEE) received the PhD degree from the Sapienza University of Rome, Italy, in 2009. After his PhD, he was a postdoc researcher with Vrije Universiteit Amsterdam, The Netherlands. He is currently a full professor with the University of Padua, Italy, and a Wallenberg-WASP guest professor with Örebro University. He has been a visiting researcher with GMU, UCLA, UCI, TU Darmstadt, UF, and FIU. His research focuses on security and privacy, where he has published more than 550 papers in international peer-reviewed journals and conferences.
\end{IEEEbiographynophoto}




\end{document}